\documentclass{pnastwo-mod2}

\usepackage{amssymb,amsfonts,amsmath}

\usepackage{ulem}
\usepackage{graphicx}

\url{www.pnas.org/cgi/doi/10.1073/pnas.1409552111}
\copyrightyear{2014}
\issuedate{}
\volume{}
\issuenumber{}

\begin{document}

\title{Spiral precipitation patterns in confined chemical gardens}

\author{Florence Haudin\affil{1}{Nonlinear Physical Chemistry Unit, Facult\'e des Sciences, Universit\'e Libre de Bruxelles (ULB), CP231, 1050 Brussels, Belgium},
Julyan H. E. Cartwright\affil{2}{ Instituto Andaluz de Ciencias de la Tierra, CSIC-Universidad de Granada, Campus Fuentenueva, E-18071 Granada, Spain.},
Fabian Brau\affil{1}{}\and A. De Wit\affil{1}{}}

\contributor{Submitted to Proceedings of the National Academy of Sciences
of the United States of America}


\maketitle

\begin{article}
\begin{abstract}
{Chemical gardens are mineral aggregates that grow in three dimensions with plant-like forms and share properties with self-assembled structures like nano\-scale tubes, brinicles or chimneys at hydrothermal vents. The analysis of their shapes remains a challenge, as their growth is influenced by osmosis, buoyancy and reaction-diffusion processes. Here we show that chemical gardens grown by injection of one reactant into the other in confined conditions feature a wealth of new patterns including spirals, flowers, and filaments. The confinement decreases the influence of buoyancy, reduces the spatial degrees of freedom and allows analysis of the patterns by tools classically used to analyze two-dimensional patterns. Injection moreover allows the study in controlled conditions of the effects of variable concentrations on the selected morphology. We illustrate these innovative aspects by characterizing quantitatively, with a simple geometrical model, a new class of self-similar logarithmic spirals observed in a large zone of the parameter space.}
\end{abstract}



\dropcap{C}hemical gardens, discovered more than three centuries ago~\cite{Glauber}, are attracting nowadays increasing interest in disciplines as varied as chemistry, physics, nonlinear dynamics and materials science. Indeed, they exhibit rich chemical, magnetic and electrical properties due to the steep pH and electrochemical gradients established across their walls during their growth process~\cite{ruiz12}. Moreover, they share common properties with structures ranging from nanoscale tubes in cement~\cite{cement_hydration}, corrosion filaments~\cite{corrosion} to larger-scale brinicles~\cite{brinicles} or chimneys at hydrothermal vents~\cite{corl79}. This explains their success as prototypes to grow complex compartmentalized or layered self-organized materials, as chemical motors, as fuel cells, in microfluidics, as catalysts, and to study the origin of life~\cite{life,life2,energy_storage,microtubes,Thouvenel_Romans2003,motors,paga08,makki09,makki12,makki12b,rosz13,makki14}. However, despite numerous experimental studies, understanding the properties of the wide variety of possible spatial structures and developing theoretical models of their growth remains a challenge.

In 3D systems, only a qualitative basic picture for the formation of these structures is known. Precipitates are typically produced when a solid metal salt seed dissolves in a solution containing anions such as silicate. Initially, a semi-permeable membrane forms, across which water is pumped by osmosis from the outer solution into the metal salt solution, further dissolving the salt. Above some internal pressure, the membrane breaks, and a buoyant jet of the generally less dense inner solution then rises and further precipitates in the outer solution, producing a collection of mineral shapes that resembles a garden. The growth of chemical gardens is thus driven in 3D by a complex coupling between osmotic, buoyancy and reaction--diffusion processes \cite{Coatman1980,Cartwright2002}. 

Studies have attempted to generate reproducible micro- and nano-tubes by reducing the erratic nature of the 3D growth of chemical gardens~\cite{microtubes,Thouvenel_Romans2003,paga08,makki12,optical_control}. They have for instance been studied in microgravity to suppress buoyancy~\cite{Jones1998,Cartwright2011b}, or by injecting aqueous solutions of metallic salts directly into silicate solutions in 3D to dominate osmotic processes by controlled flows~\cite{microtubes,Thouvenel_Romans2003}. Analysis of their microstructure has also been done for different metallic salts, showing a difference of chemical composition on the inner and the outer tube surfaces \cite{paga07bis,Cartwright2011a}. The experimental characterization and modeling of the dynamics remains however dauntingly complex in 3D, which explains why progress in quantitative analysis remains so scarce.

We show here that growing chemical gardens in a confined quasi-2D geometry by injecting one reagent solution into the other provides a new path to discover numerous original patterns, characterize quantitatively their properties and explain their growth mechanism. A large variety of structures including spirals, filaments, worms, and flowers is obtained in a horizontal confined geometry when varying the reagent concentrations at a fixed flow rate. The patterns differ from those in 3D as the growth methodology decouples the different effects involved in the formation of classical chemical gardens. The buoyancy force is reduced by the vertical confinement while injection decreases the influence of osmotic effects. 
\section{Experimental results}
\begin{figure*}
\centerline{\includegraphics[width=\textwidth]{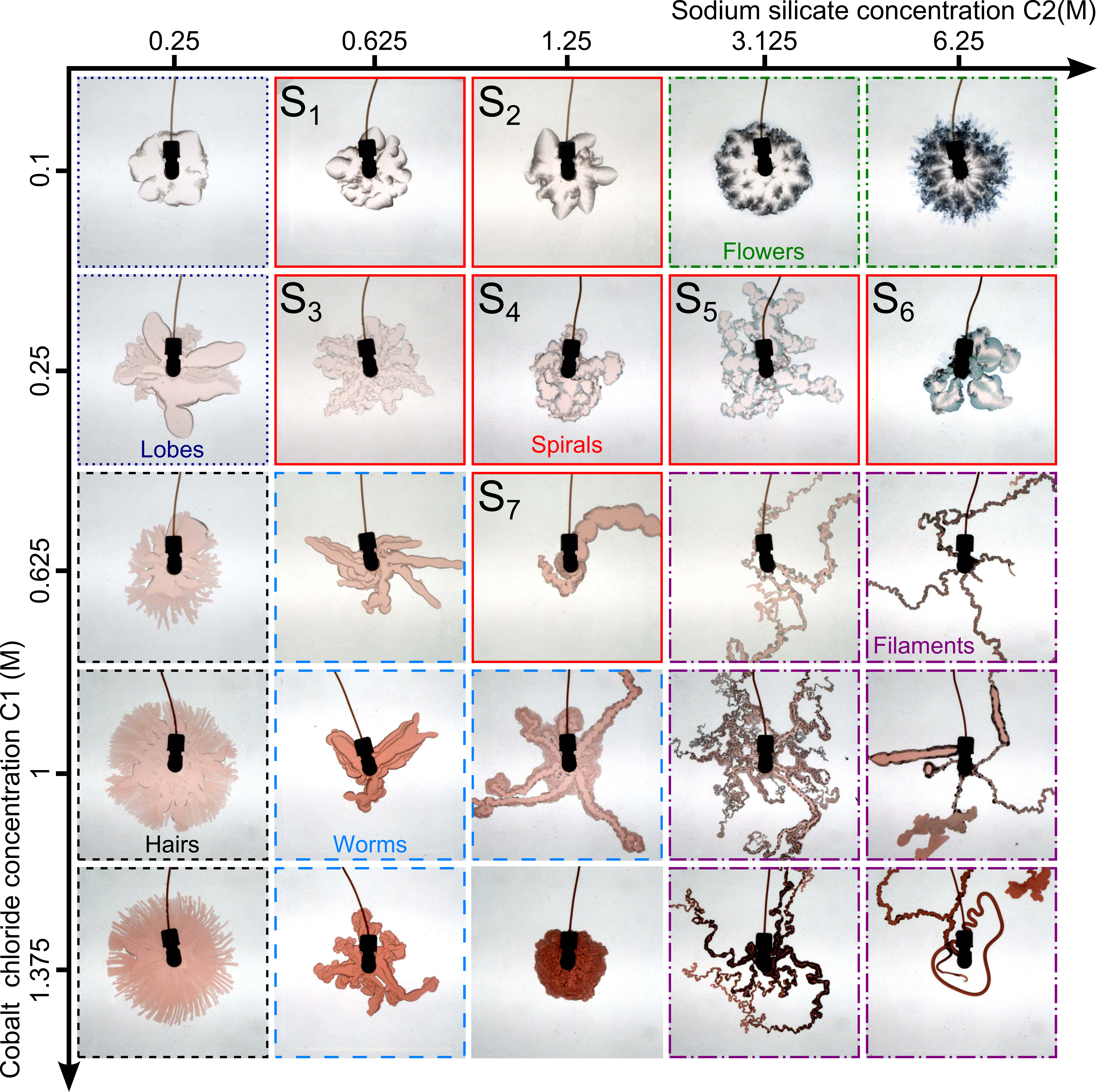}}
\caption{\textbf{Experimental patterns}. Classification of confined gardens in a parameter space spanned by the concentration $C_1$ of the injected aqueous solution of cobalt chloride and the concentration $C_2$ of the displaced aqueous solution of sodium silicate. The
diagram is divided in different colored frames referring to the various classes of patterns observed: lobes (dotted dark blue), spirals (solid red), hairs (dashed black), flowers (dashed-dotted green), filaments (long dashed-dotted purple) and worms (long dashed blue). The separation between each domain is not sharp as, for example, worms are sometimes delimited by curly boundaries that are reminiscent of spirals. The spiral category is divided into sub-categories $S_i$ analyzed in Fig.\ref{spiral} and in the SI Text. The injection rate is $Q=$ 0.11 mL/s and the field of view is 15 cm $\times$ 15 cm, shown 15~s after injection starts. Movie S1 shows some of the dynamics.}
\label{diagram}
\end{figure*}

Experiments are conducted in a horizontal Hele-Shaw cell consisting in two transparent acrylate plates separated by a small gap initially filled by a solution of sodium silicate. A solution of cobalt chloride (CoCl$_2$) is injected radially from the center of the lower plate at a fixed flow rate. Upon contact and displacement of one reactant solution by the other one, precipitation occurs and various dynamics and patterns are observed when the concentration of sodium silicate and CoCl$_2$ are varied (see Fig.\ref{diagram}).

The global trend of the phase diagram is that, if one reagent is much more concentrated than the other one, a rather circular precipitation pattern is obtained. This precipitate is concentrated at the outer rim as dark petals of {\it flowers} out of which viscous fingers grow if the sodium silicate is the more concentrated reagent. At the beginning of the injection, precipitation occurs inside the viscous fingers whereas at longer times, the precipitate lags behind them. If, on the contrary, the metal salt is much more concentrated than the sodium silicate, a compact circular pink precipitate grows radially. Above a critical radius, we observe a destabilization of the circle rim towards small scale {\it hairs} growing radially with a characteristic wavelength. 

When the concentration of both reactants is in the large range of Fig.\ref{diagram}, thin {\it filaments} growing with complex turn arounds are observed. Along the fifth column of the phase diagram, the transition from flowers to filaments occurs smoothly when the concentration of CoCl$_2$ is increased at this fixed large concentration in sodium silicate. 

Intermediate concentrations are characterized by {\it worms} as seen in the lower middle zone of the diagram. These worm patterns are reminiscent of structures observed in micellar systems when a viscous gel product is formed {\it in-situ} at the interface between reactive aqueous solutions~\cite{pod07}. At later time, terrace-like precipitate layers grow along the first structure. 

Along the lower line of the phase diagram, {\it i.e.} at a large concentration in CoCl$_2$, the transition upon an increase of the sodium silicate concentration from hairs to worms and eventually filaments transits at intermediate values via the most compact precipitate structure from all those observed. 

\begin{figure}
\centerline{\includegraphics[width=\columnwidth]{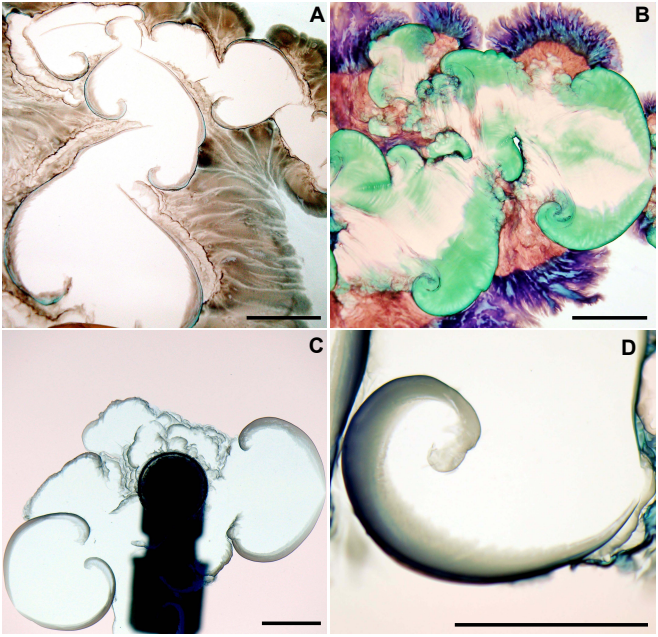}}
\caption{\textbf{Spiralling precipitates}. Panels {\bf A} and {\bf B} feature spirals and the subsequent precipitates with various colors growing diffusively out of the semi-permeable spiral walls when cobalt chloride is injected into sodium silicate ({\bf A}: case $S_3$ and {\bf B}: case $S_7$ of Fig.~\ref{diagram}). Pictures are taken a few minutes after the end of the injection. Panels {\bf C} and {\bf D} shows examples of spirals in a reverse chemical garden obtained when sodium silicate in concentration 0.625 M is injected into pink cobalt chloride 0.25 M (inverse of case S$_3$, \textbf{C}: $t=24$ s, $Q=0.03$ mL/s and \textbf{D}: few minutes after the end of injection, $Q=0.11$ mL/s). The scale bars correspond to 1 cm.}
\label{spiral_examples}
\end{figure}

In the upper middle zone (squares $S_1$ to $S_7$ in Fig.\ref{diagram}), {\it spirals} growing upon successive break-ups of precipitate walls are observed. A zoom in on some of these spirals (Fig.\ref{spiral_examples}) shows that, after a longer time, other precipitates with a rich variety of colors start growing diffusively out of their semi-permeable walls. A preliminary test with Raman microscopy has shown that some green precipitates (like in Fig.\ref{spiral_examples} for instance) is made of Co(OH)$_2$ and Co$_3$O$_4$, compounds also found in the inner surface of 3D tubes~\cite{Cartwright2011a}. This highlights the fact that different types of solid phases can form in the reaction zone between the two solutions~\cite{bon}. 

We note that friction phenomena with the cell plates do not seem to be crucial to understand the properties of the patterns as no stick/slip phenomena are observed during the growth of the patterns. Preliminary data also show that the area $A$ enclosed by the visible spiral, worm and filament patterns scales linearly with time with an angular coefficient proportional to the flow rate $Q$ divided by the gap $b$ of the Hele-Shaw cell, {\it i.e.} $A=Q t/b$. This suggests that the related patterns hence span the whole gap of the reactor. Additional experiments must be conducted to obtain statistical information on such growth properties for these patterns and for the other structures observed. 

In the following, we will focus on the spiral shape precipitate, since it is a robust pattern existing over a quite large range of concentrations and is also formed in reversed gardens, {\it i.e.} when sodium silicate is used as the injected fluid (Fig.\ref{spiral_examples}C and D). 

\section{Geometrical model}
\begin{figure*}
\centerline{\includegraphics[width=0.8\textwidth]{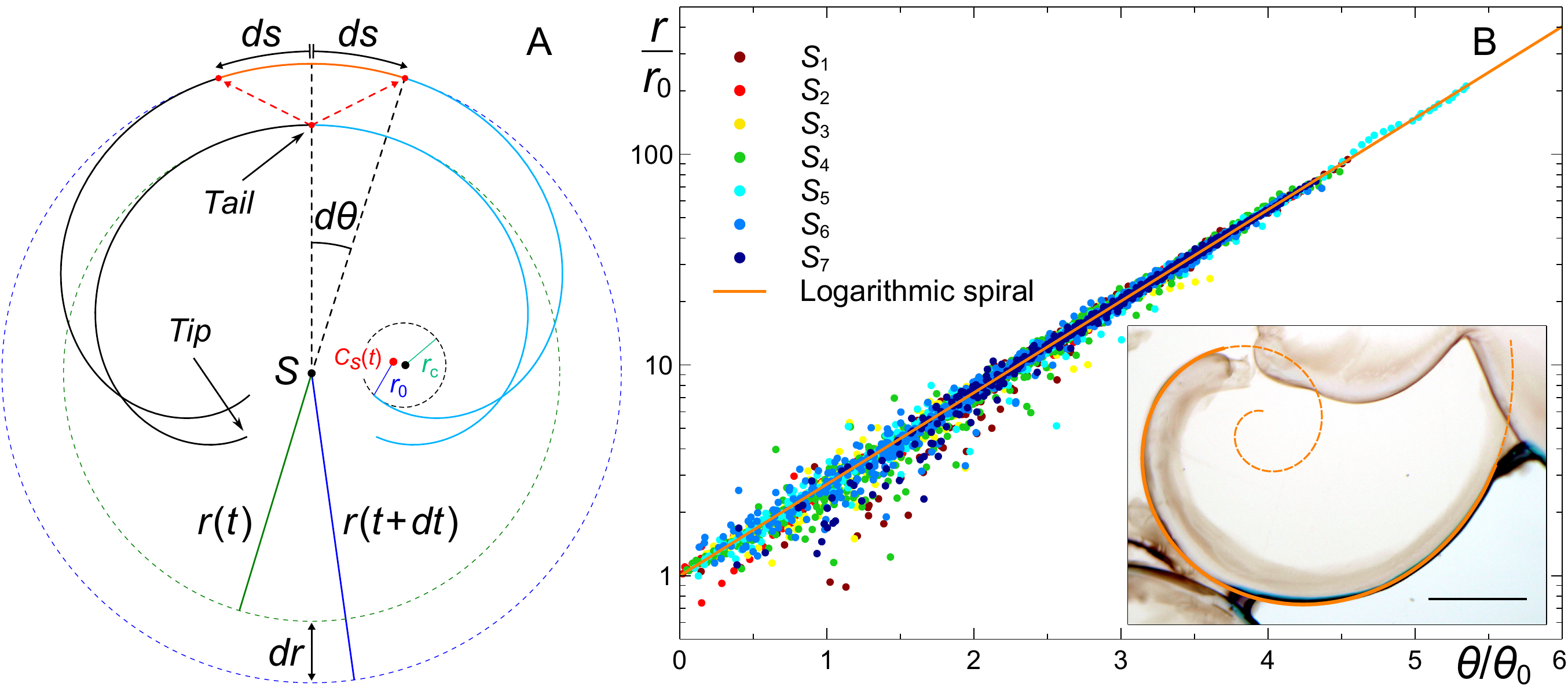}}
\caption{\textbf{Spiral growth mechanism and scaling of logarithmic spiral-shaped precipitates.} {\bf A,} Schematic of the growth mechanism of the curly-shaped precipitates during an infinitesimal interval of time where $r(t)$ is the radius of the expanding bubble of injected reagent and $S$ is the point source. {\bf B,} Scaling law for the experimentally measured evolution of the radial distance $r/r_0$ as a function of the scaled angle $\theta/\theta_0$ for 173 spirals (9 experiments) corresponding to the seven different categories $S_i$ in Fig.~\protect\ref{diagram}. The inset shows a superimposition of a logarithmic spiral on a spiraled precipitate (the scale bar corresponds to 2 mm). The procedures for the measurements of $r(\theta)$ and the distributions of $r_0$ and $\theta_0$ are given in the SI Text.}
\label{spiral}
\end{figure*}

The spiral formation mechanism can be understood with a minimal geometrical argument. Consider a reagent 1 injected radially from a source point $S$ into reagent 2 in a 2D system (Fig.\ref{spiral}A). The contact zone between the reactive solutions, in which precipitation occurs, initially grows as a circle of arbitrarily small radius $r$. If, because of further injection, this layer of precipitate breaks at a critical value $r=r_{\rm{c}}$, the new precipitate, formed as the bubble of reagent 1 expands, pushes the already existing solid layer. As a result, the branch of solid precipitate starts being advected out of the growing bubble and rotates as a whole around the breaking point which later on identifies to being the tip of the spiral (see also Movies S2 and S3 of growing spirals). An arc of spiral is hence observed to develop with its tip moving in the fixed frame of reference centered on $S$. It further grows by precipitation at its tail located on the circle of growing radius $r(t)$. 

The equation of the resulting curly-shaped precipitation layer is obtained by considering two infinitesimally close time steps as shown in Fig.\ref{spiral}A. At time $t+dt$, the precipitate layer formed at time $t$ is pushed away by the newly added material and rotates by an angle $d\theta$. The precipitate layer produced between $t$ and $t+dt$ is created from the reaction occurring at the new section of the contact line between the two reagents~\cite{douady}. Since this new section of contact line is generated by the bubble expansion, its length $ds$ is proportional to the increase of perimeter ($\sim dr$) of the expanding bubble, namely
\begin{equation}
\label{ds1}
ds =\theta_0 \ dr(t)
\end{equation}
where $\theta_0$ is a constant growth rate controlling the length of the precipitate layer created as the bubble radius increases. A large value of $\theta_0$ implies that the length $ds$ of the new segment of precipitate created at $t+dt$ is large compared to the increase of length $dr$ along the radial direction and leads to a more coiled spiral. 

Upon integration of Eq.(\ref{ds1}) we get $s =\theta_0 (r-r_{\rm{c}})$, where $s$ is the length of the curve from the tip to the tail. The constant of integration was fixed by considering that the tip of the curve ($s=0$) is generated from a circle of radius $r_{\rm{c}}$. In addition, the radius $r$ of the expanding bubble coincides with the radius of curvature $R(s)$ of the tail since the bubble is also the osculating circle (see Fig.\ref{spiral}A). Therefore, we obtain $R(s)=r_{\rm{c}}+ s/\theta_0$ which is the Ces\`aro equation of a logarithmic spiral giving the evolution of the radius of curvature along the curve as a function of the arclength~\cite[p. 26]{stru88} (see SI Text for more details).

Alternatively, we can also proceed as follows to obtain the expression of the spiral curve in polar coordinates. To derive this equation, the relation existing between $dr$ and $d\theta$ should be known everywhere along the curve in the fixed frame of reference centered around $S$. For non moving curves, this relation is usually found by analyzing how $dr$ varies in a sector of constant central angle $d\theta$ which rotates to span the entire curve. In our case, the growing arc of spiral is moving in a fixed system of coordinates centered on the source point $S$. Its $r(\theta)$ equation is therefore alternatively obtained by considering a non moving sector of constant central angle $d\theta$ as in Fig.\ref{spiral}A and analyzing how $dr$ varies in it when the curve spans this angular sector thanks to its motion (see SI Text).

As seen in Fig.\ref{spiral}A, the length $ds$ of the precipitate layer added at $t+dt$ in the fixed sector of angle $d\theta$ is given by $ds=r(t+dt) d\theta$ which, at first order in $dt$ and $d\theta$, reduces to 
\begin{equation}
\label{ds2}
ds= r(t) d\theta.
\end{equation}
Combining (\ref{ds1}) and (\ref{ds2}), we get 
\begin{equation}
\label{drdtet}
d\theta = \theta_0 \frac{dr}{r}
\end{equation}
which is readily integrated to yield
\begin{equation}
r=r_0\, e^{\theta/\theta_0}.
\label{log}
\end{equation}
$r_0$ is a constant of integration that can be computed for each spiral if we note that $\theta=0$ fixes the starting point of the spiral {\it i.e}. its tip. Reminding that the tip originated from the breaking by injection of the initial small circle of radius $r_{\rm{c}}$, the radius of curvature of the spiral at $\theta=0$ is nothing else than $r_{\rm{c}}$. Therefore, $r_{\rm{c}} = r_0 \sqrt{1+\theta_0^2}/\theta_0$ which relates the constant of integration $r_0$ to $r_{\rm{c}}$. It is of interest to note that Eq.(\ref{log}) exactly describes the (blue) segments of spiral of Fig.\ref{spiral}A in the frame of reference centered on $S$ provided this segment is translated and rotated such that the center $C_{\rm{S}}$ of the spiral coincides with $S$ (see SI Text). 

Equation (\ref{log}) constructs a logarithmic spiral in polar coordinates and quantitatively describe structures in many natural systems such as seashells, snails, or the horns of animals where the growth mechanism preserves the overall shape by the simple addition of new material in successive self-similar steps~\cite{thom92}. It appears thus logical that a similar logarithmic shape is recovered here in the case of regular additional precipitation at the tail of a growing arc of solid precipitate upon further injection of reactants at a fixed flow rate. Equation (\ref{log}) quantitatively describes the spiral structures observed in a large part of our phase diagram as demonstrated in the next section. 

\section{Comparison between experiments and model}

To test our geometrical model, the radii of 173 spirals observed in 9 experiments for 7 pairs of concentrations (sectors $S_i$ of Fig.\ref{diagram}) have been measured as a function of the polar angle (see Fig.\ref{spiral}B and SI Text). We have selected precipitate layers which were not deformed by other structures growing in their neighborhood or fractured. Among them, we have also selected sufficiently coiled segments such that their polar angle span the interval $[0,\theta_{\rm{max}}]$ with $\theta_{\rm{max}} > 2.44$ $(140^\circ)$ in order to obtain significant constraints on the model, \textit{i.e.} $\theta / \theta_0$ large enough in Fig.3B. Indeed, segments which are not sufficiently coiled cannot be used to determine the type of spiral emerging in the system. If the interval spanned by $\theta$ is too small, Eq.(4) is hardly distinguishable from the equation $r=r_0 (1+ \theta/\theta_0)$ describing an archimedean spiral. The distribution of $\theta_{\rm{max}}$ among the analyzed spirals is shown in Fig.S7. As seen in Fig.\ref{spiral}B, where the radial distance and the polar angle are rescaled by $r_0$ and $\theta_0$ respectively, all the analyzed spiral profiles collapse onto an exponential master curve, illustrating that indeed they are all logarithmic to a good accuracy.

The dispersion occurring at low $\theta$ (near the spiral center) is related to the fact that the experimental spiral structures emerge from an arc of an initial tiny circular section of radius $r_{\rm{c}}$. 63 other spirals corresponding to the inverted cases $S_1$ and $S_3$ ({\it i.e.} sodium silicate injected into cobalt chloride) have also been analyzed and prove to follow the same logarithmic scaling law (see Fig.S4H). We note that, in our confined gardens, the spiral growth continues only as long as the precipitate can pivot within the system. When it becomes pinned by encountering a solid wall or another precipitate, the spiral growth ceases, the membrane breaks, and a new radial source is produced, leading regularly to a new fresh spiral. This behavior is reminiscent of the periodic pressure oscillations reported in the growth of some 3D chemical gardens~\cite{paga08,Pantaleone09}. 

Thus self-similar logarithmic spirals emerging in a significant part of the phase diagram can be quantitatively analyzed on the basis of a simple geometrical argument. The model, which describes an isolated source producing a circular precipitation zone that then ruptures, proves to be robust even when spirals are interacting, as in the current experiments.

\section{Conclusions}
This work shows that new insight into the complexity of self-assembled chemical garden structures can be obtained by growing them by injection of reactive solutions in confined geometries. The control of the concentrations of the reagents and of the flow rate will allow the study of phase diagrams in reproducible experimental conditions~\cite{makki12,toth07}, as well as the switch from dominant reaction--diffusion processes to flow-driven ones. Moreover, the quasi-2D nature of the precipitates will permit, as done here, an easier characterization of the patterns using tools of classical 2D pattern selection analysis~\cite{2Dpatterns}. Modeling and numerical simulations of this injection-driven aggregation process will thus be simplified. This will facilitate the analysis of growth mechanisms and of the relative effects of reactions, hydrodynamics and mechanics in the resulting structure; a prerequisite to the rational design of complex, hierarchical microarchitectures~\cite{microtubes,noor13}.

\begin{materials}
The experimental set-up is a horizontal Hele-Shaw cell consisting of two transparent acrylate plates of size 21.5 cm x 21.5 cm x 0.8 cm separated by a gap of 0.5 mm and initially filled with a sodium silicate solution of concentration $C_2$. A cobalt chloride solution of concentration $C_1$ is injected into the sodium silicate solution of concentration $C_2$. Acrylate was chosen instead of glass to avoid interactions between the sodium silicate and the glass. For visualization purposes, the cell is placed on top of a diffuse light table and the dynamics is recorded from above using either a CMOS camera or a photography camera.
Some earlier work on chemical gardens employed a similar confined geometry but used seed crystals rather than injection which leads to different dynamics than those studied here~\cite{Cartwright2002}. The metallic salt solution is prepared from dissolution in water of crystals of cobalt (II) chloride hexahydrate (CoCl$_2\cdot$6H$_2$O) (Sigma-Aldrich). The sodium silicate solution is prepared from a commercial aqueous solution (Sigma-Aldrich), with formula Na$_2$(SiO$_2$)$_x$ $\times$ H$_2$O and the composition SiO$_2$ $\approx$ 26.5\% and Na$_2$O $\approx$ 10.6\%. Cobalt chloride is one of the classical metallic salts used to grow chemical gardens in 3D: structures form with a smaller induction time and a higher linear growth than with calcium, nickel and manganese salts, for instance~\cite{Cartwright2011a}. In Fig.\ref{diagram}, we used a constant injection rate of 0.11 mL/s, which is larger than those typically used for 3D injection experiments~\cite{Thouvenel_Romans2003} and leads to a growth process mainly driven by the flow. We vary the concentration of sodium silicate within the same range as in 3D experiments~\cite{Cartwright2011a}, from 0.25 to 6.25~M. The cobalt chloride concentration is varied from 0.100 to 1.375~M.
\end{materials}

\begin{acknowledgments}
 We thank P. Borckmans and C.I. Sainz-D\'{\i}az for discussions. We are very grateful to K. Baert and I. Vandendael from Vrije Universiteit Brussel (VUB) for the Raman analysis. A.D., F.B. and F.H. acknowledge PRODEX and FRS-FNRS (FORECAST project) for financial support. J.H.E.C. acknowledges the financial support of MICINN grant FIS2013-48444-C2-2-P.

\end{acknowledgments}

\newpage

\noindent {\Large {\bf Supporting Information}}

\section{Geometrical model: Alternative derivation}

\subsection{Intrinsic equation}

We give here an alternative and equivalent derivation of the equation describing the curve generated by the geometrical model. This derivation is independent on the coordinate system and leads to a so-called natural (or intrinsic) equation giving the evolution of the local radius of curvature of the curve as a function of the arclength~\cite[p. 26]{stru88}. 

We still consider the system at two infinitesimally close time steps such as the length of the curve has increased by an amount $ds$ given by Eq.(1) of the main text, namely
\begin{equation}
\label{ds}
ds=\theta_0\, dr_{\rm{b}}= \theta_0 \frac{dr_{\rm{b}}}{dt}dt,
\end{equation}
where $r_{\rm{b}}(t)$ is the radius of the bubble of injected reagent at time $t$ (see Fig.\ref{fig-schema}A). The curve is parametrized by its arclength $s$ such that the value of $s$ of an arbitrary point $P$ is equal to the length of the curve measured from the first point $P_0$ to $P$ (see Fig.\ref{fig-schema}A). As explained in the main text, $P_0$ is generated from an initial small circle of radius $r_{\rm{b}}(0)=r_{\rm{c}}$. The radius of curvature at $s=0$ is thus given by $R(0)=r_{\rm{c}}$. At time $t$, the radius of curvature of the last point of the curve at its tail is given by
\begin{equation}
\label{rt}
R(s)=r_{\rm{b}}(t).
\end{equation}
Similarly, at time $t+dt$, the radius of curvature of the last point of the curve is given by
\begin{equation}
\label{rtdt}
R(s+ds)=r_{\rm{b}}(t+dt)=r_{\rm{b}}(t)+\frac{dr_{\rm{b}}}{dt}dt,
\end{equation}
where we used a first order expansion in $dt$. Therefore, the difference between Eq.(\ref{rtdt}) and Eq.(\ref{rt}) leads to
\begin{equation}
R(s+ds)-R(s) = \frac{dR}{ds} ds = \frac{dr_{\rm{b}}}{dt}dt,
\end{equation}
where we used a first order expansion in $ds$. Using the expression (\ref{ds}) of $ds$, we obtain
\begin{equation}
\frac{dR}{ds} = \frac{1}{\theta_0}.
\end{equation}
Integration of this last equation leads to
\begin{equation}
R(s)=R(0) + \frac{s}{\theta_0} = r_{\rm{c}} + \frac{s}{\theta_0}.
\end{equation}
The radius of curvature is thus a linear function of the arclength. This is the Ces\`aro equation of a logarithmic (equiangular) spiral~\cite[p. 26]{stru88}. Let us now obtain the equation of the curve in polar coordinates.

\subsection{Polar equation}

We have thus obtained above the curvature $\kappa$ as a function of the arclength $s$:
\begin{equation}
\kappa(s) = \frac{1}{R(s)} = \frac{\theta_0}{\theta_0 r_{\rm{c}} +s}.
\end{equation}
The relation between the Ces\`aro equation and the Cartesian coordinates is~\cite[p. 26]{stru88}
\begin{align}
\label{ces-cart}
x(s) &= \int \cos \bar{\kappa}(s) ds= \frac{\theta_0 r_{\rm{c}} +s}{1+\theta_0^2}(\cos \bar{\kappa}(s) + \theta_0 \sin \bar{\kappa}(s) ), \\
y(s) &= \int \sin \bar{\kappa}(s) ds= \frac{\theta_0 r_{\rm{c}} +s}{1+\theta_0^2}(\sin \bar{\kappa}(s) - \theta_0 \cos \bar{\kappa}(s) ),
\end{align}
where,
\begin{equation}
\label{kappa-bar}
\bar{\kappa}(s) = \int \kappa(s) ds = \theta_0 \ln(\theta_0 r_{\rm{c}} +s) + K,
\end{equation}
and $K$ is a constant of integration which fixes the orientation of the curve in space and is fixed below. The constants of integration in Eq.(\ref{ces-cart}) correspond to translations of the curve and are set to zero.

Using Eqs.~(\ref{ces-cart}) and (\ref{kappa-bar}), the polar equation is obtained as follows
\begin{equation}
\label{radius-s}
r(s)= \sqrt{x^2(s)+y^2(s)}=\frac{\theta_0 r_{\rm{c}} +s}{\sqrt{1+\theta_0^2}},
\end{equation}
and
\begin{eqnarray}
\label{theta-s}
\theta(s) &=& \arctan\left(\frac{y}{x}\right)= \arctan\left(\frac{\sin \bar{\kappa}(s) - \theta_0 \cos \bar{\kappa}(s)}{\cos \bar{\kappa}(s) + \theta_0 \sin \bar{\kappa}(s)}\right) \nonumber \\
&=& \arctan\left(\frac{\tan \bar{\kappa}(s)-\theta_0}{1+\theta_0\tan \bar{\kappa}(s)}\right) \nonumber \\ 
&=& \arctan\left[\tan(\bar{\kappa}(s)-\arctan \theta_0)\right] \nonumber \\
&=& \bar{\kappa}(s)-\arctan \theta_0  \nonumber\\
&=& \theta_0 \ln(\theta_0 r_{\rm{c}} +s) + K-\arctan \theta_0.
\end{eqnarray}
Eliminating $s$ between Eqs.~(\ref{radius-s}) and (\ref{theta-s}) we have
\begin{equation}
\label{polar-temp}
r = \frac{e^{(\arctan \theta_0-K)/\theta_0}}{\sqrt{1+\theta_0^2}} e^{\theta/\theta_0}.
\end{equation}
The constant $K$ can be fixed by imposing that $s=0$ corresponds to $\theta=0$ meaning that the radius of curvature of the curve (\ref{polar-temp}) at $\theta=0$ should be equal to $r_{\rm{c}}$. This leads to $K=\arctan \theta_0 - \theta_0 \ln(\theta_0 r_{\rm{c}})$ and Eq.(\ref{polar-temp}) reduces to
\begin{equation}
\label{log2}
r = \frac{r_{\rm{c}} \theta_0}{\sqrt{1+\theta_0^2}} e^{\theta/\theta_0} \equiv r_0\, e^{\theta/\theta_0}.
\end{equation}

\section{Discrete algorithm}

\begin{figure*}
\begin{center}
\includegraphics[width=\textwidth]{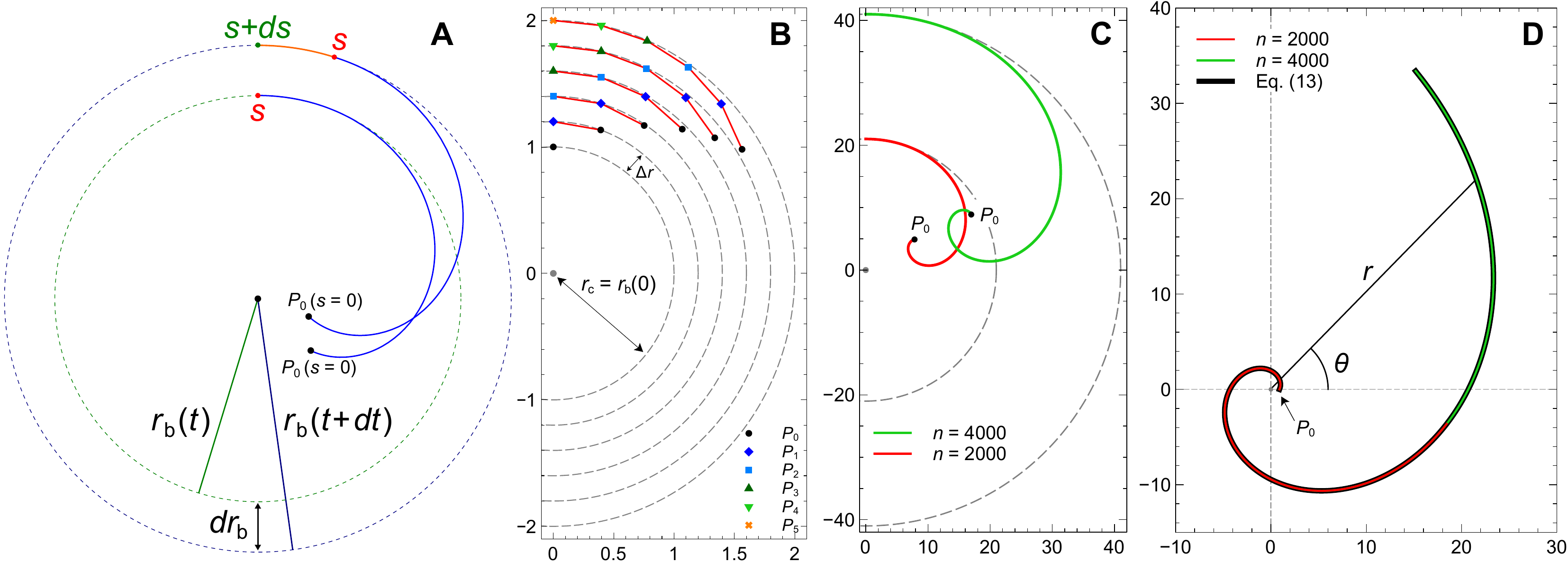}
\end{center}
\caption{{\bf A,} Schematic of the growth mechanism of the curly-shaped precipitates during an infinitesimal interval of time. {\bf B,} First few points generated by this mechanism and obtained from a discrete algorithm using $\Delta r = 0.2$, $\theta_0=2$ and $r_{\rm{c}} = 1$. The red straight lines are added to help visualize the emerging structure. Due to symmetry, the generation of only one curve is shown. {\bf C,} Curves generated from a discrete algorithm using $\Delta r = 0.01$, $\theta_0=2$ and $r_{\rm{c}} = 1$. The curves obtained after $n=2000$ and $n=4000$ iterations are shown and correspond to a bubble of injected reagent having a radius of $21 r_{\rm{c}}$ and $41 r_{\rm{c}}$ respectively ($n \Delta r + r_{\rm{c}}$). {\bf D,} If the emerging curves are logarithmic spirals, the position of the spiral center can be deduced from the position of the center of curvature of $P_0$ (see section ``Spiral analysis" below). In the context of this algorithm, the position of the center of curvature of $P_0$ is tracked during the growth process. The total angle of rotation of the curves during the growth is simply equals to the sum of all $\Delta \theta$ applied. Therefore, these two curves can be properly rotated and translated such that their centers coincide with the origin of coordinates. These curves are exactly described by Eq.(\ref{log}) with $\theta_0=2$ and $r_0= r_{\rm{c}} \theta_0/(1+\theta_0^2)^{1/2}\simeq 0.89$, such as its radius of curvature at $P_0$ ($\theta = 0$) is equal to $r_{\rm{c}}$.} 
\label{fig-schema}
\end{figure*}

The curves generated by the proposed mechanism and depicted schematically in Fig.\ref{fig-schema}A can also be directly constructed using the following discrete algorithm (Fig.\ref{fig-schema}B). We start from a circle of radius $r_{\rm{c}}$ with a point $P_0$ at the top which represents the first point of generated curve. 
\begin{enumerate}
\item The radius of the circle is increased by a given $\Delta r$. The ordinate of $P_0$ is increased by $\Delta r$ and then rotated by an angle $\Delta \theta$. A point $P_1$ is added at the top of the circle and represents a new amount of precipitate.
\item The radius of the circle is increased by a constant $\Delta r$. The ordinates of $P_0$ and $P_1$ are increased by $\Delta r$ and then are both rotated by an angle $\Delta \theta$. A point $P_2$ is added at the top of the circle.
\end{enumerate}
The procedure is then iterated to generate the next points. The discrete angle used is provided by Eq.(3) of the main text: $\Delta \theta = \theta_0 \Delta r /r$. Figure~\ref{fig-schema}B shows the generation of the first few points using a large value of $\Delta r$ for clarity. Figure~\ref{fig-schema}C shows the resulting structures emerging from the 2000 and 4000 iterations of the algorithm using a smaller value of $\Delta r$. Figure~\ref{fig-schema}D shows that these two curves, once properly rotated and translated, are exactly described by Eq.(\ref{log}) using the same value of $\theta_0$ and a value of $r_0$ such that the radius of curvature at $P_0$ is equal to the radius $r_{\rm{c}}$ of the initial small circle, namely $r_0= r_{\rm{c}} \theta_0/(1+\theta_0^2)^{1/2}$.

\section{Spiral analysis}
\label{sec:spi-anal}

Each experiment is recorded by taking photos at regular time steps adapted according to the injection rate. Typically the time interval between two photos is 1 s for the injection rates used. The pattern is then analyzed at one given time for each experiment when the spirals are sufficiently developed and/or when the overall pattern is as large as the field of view. To analyze the spiral observed in our experiments, we measure the evolution of the spiral radii, $r$, as a function of the polar angle $\theta$. By convention, the starting point $P_0$ of the spiral, which is the closest to the spiral center, is characterized by $\theta=0$ and $r=r_0$ (Fig.\ref{fig01}). However, the exact position of the spiral center, $C_{\rm{S}}$, from which the radii should ideally be measured, is not known. In this section, we explain the method used to overcome this difficulty.

\subsection{System of coordinates centered on the spiral center}

The equation of a logarithmic spiral written in a system of coordinates centered on the spiral center, $C_{\rm{S}}$, is given by 
\begin{equation}
\label{log-spiral}
r=r_0 e^{\theta/\theta_0},
\end{equation}
where $r_0$ and $\theta_0$ are constant parameters. If the spiral radii, $r$, are measured from the exact position of the spiral center, $C_{\rm{S}}$, then the evolution of $r$ as a function of the polar angle $\theta$ can be fitted using Eq.(\ref{log-spiral}). The measured spiral radii and the polar angle can then be rescaled by $r_0$ and $\theta_0$ respectively to produce the graph displayed in Fig.3B of the main text.

\subsection{Arbitrary system of coordinates}

The exact position of the spiral center is generally not known a priori and the radii are measured in an arbitrary system of coordinates whose origin does not coincide with $C_{\rm{S}}$. In such a system of coordinates centered on say $C_{\rm{A}}$, the expression of a logarithmic spiral is no longer given by the simple form (\ref{log-spiral}). In this section, we derive the general expression describing a logarithmic spiral off centered with regard to $C_{\rm{S}}$ which is used to fit the data.

\begin{figure}[h]
\begin{center}
\includegraphics[width=\columnwidth]{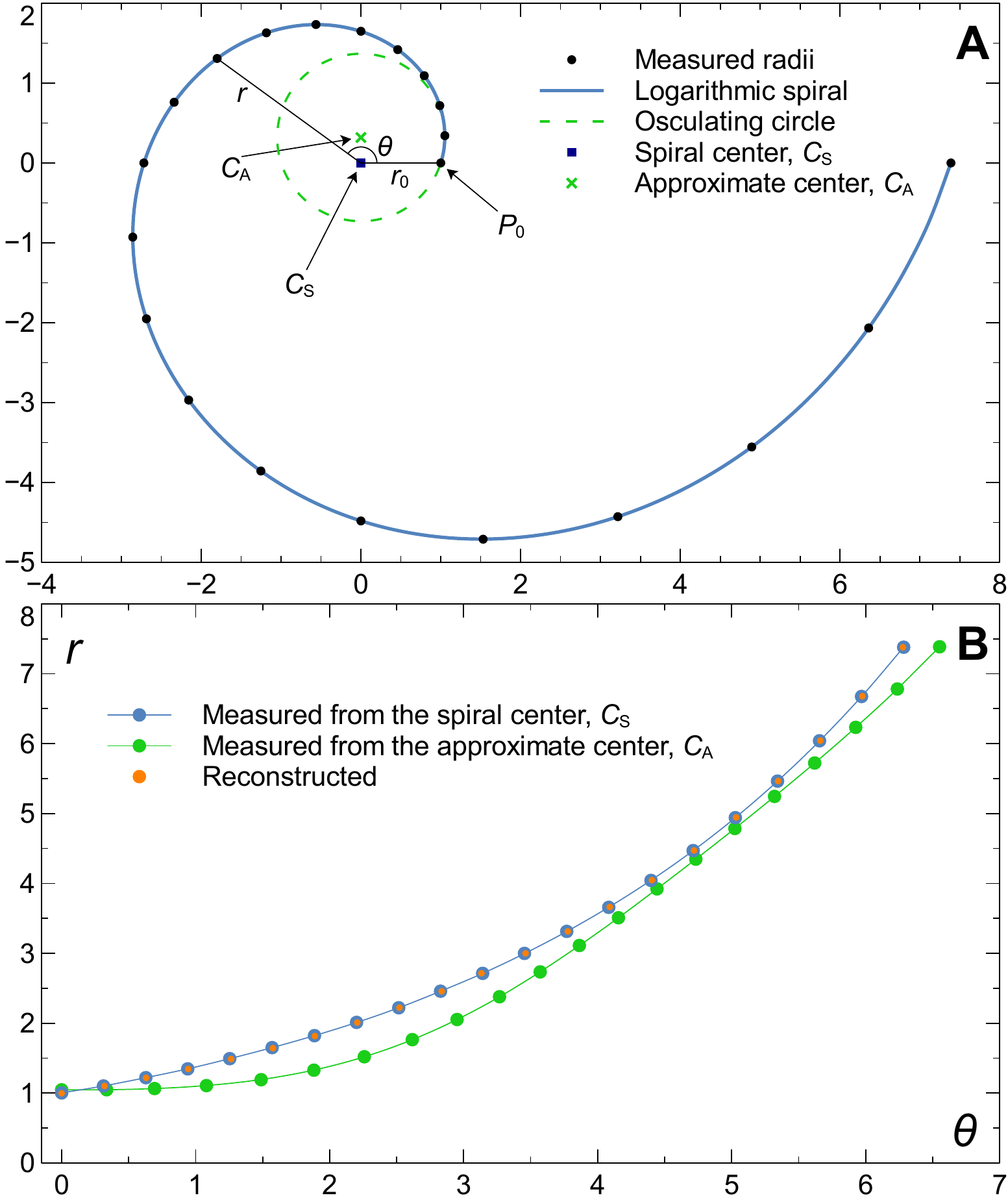}
\end{center}
\caption{{\bf A,} Logarithmic spiral together with the osculating circle passing through the point $P_0$ which is the closest to the spiral center. The center of the osculating circle, $C_{\rm{A}}$, is used as approximate center to measure the spiral radii. The black dotes show where the radii are measured. $C_{\rm{S}}$ indicates the position of the spiral center. {\bf B,} Plot of the radius $r$ of the spiral as a function of the polar angle $\theta$ using the exact and the approximate positions of the spiral center. The curve reconstructed from the data obtained with the approximate position of the spiral center is also shown (orange dots) and agrees well with the spiral curve measured directly from the exact center position $C_{\rm{S}}$.} 
\label{fig01}
\end{figure}

The approximate position of the spiral center, $C_{\rm{A}}$, used to measure the spiral radii is obtained from the osculating circle passing through $P_0$. Figure~\ref{fig01}A shows this construction on an exact logarithmic spiral to illustrate and test the proposed procedure. Figure~\ref{fig01}B shows the evolution of the spiral radii, $r$, as a function of the polar angle, $\theta$, when the radii are measured from the exact ($C_{\rm{S}}$) and approximate ($C_{\rm{A}}$) positions of the spiral center respectively. The polar coordinates of the spiral obtained from the approximate position of the center is noted $(r',\theta')$ whereas the polar coordinates of the spiral obtained from the exact position of the center is noted $(r,\theta)$. 

\begin{figure}
\begin{center}
\includegraphics[width=\columnwidth]{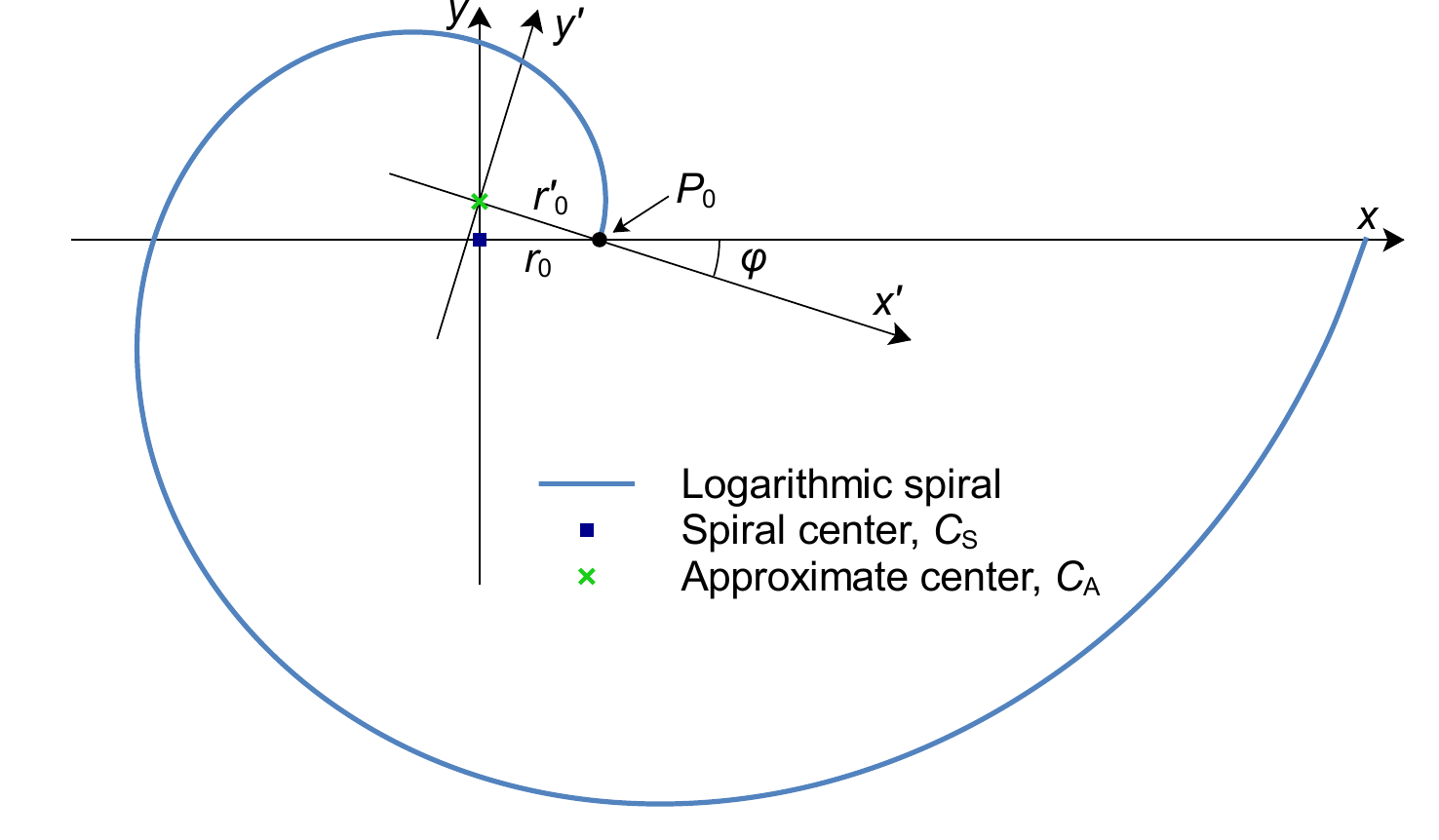} 
\end{center}
\caption{Representation of the two systems of coordinates. The system of coordinates $(x,y)$ is centered on the spiral center $C_{\rm{S}}$ whereas the systems of coordinates $(x',y')$ is centered on $C_{\rm{A}}$ and is used to measure the spiral radii.}
\label{fig02}
\end{figure}

These two curves are obviously equivalent and describe the same spiral. As shown in Fig.\ref{fig02}, they are just measured in two different systems of coordinates related by a rotation and a translation as
\begin{equation}
\label{transf1}
\left(\begin{array}{c} x \\ y \end{array} \right)= \left(\begin{array}{cc}
\cos \varphi & -\sin \varphi \\ 
\sin \varphi & \cos \varphi
\end{array} \right) \left(\begin{array}{c} x' \\ y' \end{array} \right)-\left(\begin{array}{c} x_0 \\ y_0 \end{array} \right),
\end{equation}
with
\begin{equation}
\label{phi}
\varphi=\arcsin\left(\frac{y_0}{r'_0}\right), \quad x'= r'\cos \theta', \quad y'= r'\sin \theta'.
\end{equation}
$x_0$ and $y_0$ are the Cartesian coordinates of $C_{\rm{A}}$ in the $(x,y)$ system of coordinates whose origin coincide with the spiral center $C_{\rm{S}}$. Since $r'$, $\theta'$ and $r'_0$ are the quantities measured in practice, the curve $(r,\theta)$ obtained from the exact position of the center can thus be reconstructed once $x_0$ and $y_0$ are known by using~\footnote{Notice that if the spiral is exactly logarithmic and if the approximate center is constructed using the osculating circle passing through $P_0$, then $x_0=0$ as seen on Fig.\ref{fig01}A. In practice, the spirals are obviously never exactly logarithmic and the position of the osculating circle is never perfect. Consequently, one needs to allow for a non vanishing value for $x_0$ in the procedure.}
\begin{equation}
r=\sqrt{x^2+y^2}, \quad \theta=\arccos\left(\frac{x}{r}\right).
\end{equation}
We show below how $x_0$ and $y_0$ can be obtained.

The spiral parameters, $r_0$ and $\theta_0$, together with the translation parameters of the center, $x_0$ and $y_0$, are obtained all at once by fitting the data by the general expression of a logarithmic spiral valid in an arbitrary system of coordinates. 

This general expression is simply obtained by considering the reverse of the transformation~(\ref{transf1})
\begin{equation}
\label{transf2}
\left(\begin{array}{c} x' \\ y' \end{array} \right)= \left(\begin{array}{cc}
\cos \varphi & \sin \varphi \\ 
-\sin \varphi & \cos \varphi
\end{array} \right) \left(\begin{array}{c} x+x_0 \\ y+y_0 \end{array} \right),
\end{equation}
with
\begin{equation}
x=r_0 e^{t/\theta_0} \cos t, \quad y=r_0 e^{t/\theta_0} \sin t,
\end{equation}
and $\varphi$ is given by Eq.(\ref{phi}). The parametric equations, where $t$ is the parameter, describing a logarithmic spiral in an arbitrary system of coordinates is finally given by
\begin{subequations}
\label{parametric-eq}
\begin{eqnarray}
r'(r_0,\theta_0,x_0,y_0;t)&=&\sqrt{x'^2+y'^2} \\
\theta'(r_0,\theta_0,x_0,y_0;t)&=&\arccos\left(\frac{x'}{r'}\right).
\end{eqnarray}
\end{subequations}
The values of the parameters $r_0$, $\theta_0$, $x_0$ and $y_0$ are obtained for each spiral by fitting the expression (\ref{parametric-eq}) to the measured values of $r'$ and $\theta'$ using a nonlinear regression procedure (Mathematica).

At this stage, since $x_0$ and $y_0$ are known, we apply the transformation (\ref{transf1}) to the data in order to obtain the measured curve in a system of coordinates centered on the spiral center. The result is shown in Fig.\ref{fig01}B with a very good agreement compared to the measurements performed directly in a system of coordinates centered on the spiral center. This illustrates the correctness of the procedure proposed to treat the data. Finally, the transformed data are rescaled by $r_0$ and $\theta_0$ to produce the graph displayed in Fig.3B of the main text.

\section{Results from the analysis}

The results of the analysis described in the previous section are gathered in Fig.3B of the main text. The same results are presented here separately for each categories $S_i$ defined in Fig.1 of the main text. The spirals are logarithmic in good approximation in each of the seven sectors $S_i$ of the phase diagram (Fig.\ref{fig03} below). Only the sector $S_1$ displays more dispersion for low values of $\theta/\theta_0$ but at larger values of the rescaled polar angle, the spirals follow closely the evolution of a logarithmic spiral. For information, we also show in Fig.\ref{fig03}H the results obtained for the inverted case, where sodium silicate is injected into cobalt chloride, corresponding to the sectors $S_1$ and $S_3$. Those spirals are also logarithmic.

In Fig.\ref{fig04}, we show the distributions of the values of $r_0$ and $\theta_0$ characterizing all the analyzed spirals. Those distributions show also the contributions of each sector $S_i$. Notice that those distributions have a meaning only if $r_0$ and $\theta_0$ are both independent on the reagent concentrations (or if the dependence is weak). It seems that this is roughly the case\footnote{The only exception concerns the sector $S_2$ which is characterized by larger values of $r_0$ ($r_0 \gtrsim$ 0.5 mm). However, only 7 spirals have been analyzed for this sector. Those data are not reported in Fig.\ref{fig04}A for clarity.} by inspecting the contributions of each sector $S_i$. However, the number of analyzed spirals per sector is not large enough to draw definitive conclusions. The distributions of $r_0$ and $\theta_0$ are both rather well fitted by a log-normal distribution 
\begin{equation}
\label{log-normal}
f(x; \mu, s)= \frac{\lambda}{x}\, e^{-\frac{(\ln x - \mu)^2}{2 s^2}}.
\end{equation}
The expectation value $E$ is then given by
\begin{equation}
E=e^{\mu + s^2/2},
\end{equation}
and the standard deviation $\sigma$ is
\begin{equation}
\sigma = \left(e^{s^2}-1\right)^{1/2} \, E.
\end{equation}
We find
\begin{equation}
\label{variation-r0-t0}
r_0 = (0.43 \pm 0.20)\, \rm{mm}, \quad \theta_0 = 1.67 \pm 0.52.
\end{equation}

Within our minimal geometric model, the radius of curvature of the spiral at its starting point $P_0$ should be close to the radius $r_{\rm{c}}$ of the initial circle of solid precipitate before the solid layer breaks. The radius of curvature $R$ of a logarithmic spiral is given by
\begin{equation}
R=r_0\, e^{\theta / \theta_0}\, \frac{\sqrt{1+\theta_0^2}}{\theta_0}.
\end{equation}
Consequently, at the point $P_0$ ($\theta=0$), the radius of curvature $R_{P_0}\equiv r_{\rm{c}}$ is given by
\begin{equation}
r_{\rm{c}}=r_0\, \frac{\sqrt{1+\theta_0^2}}{\theta_0}.
\end{equation}
The distribution of $r_{\rm{c}}$ in our experiments is displayed in Fig.\ref{fig05}A and also follows a log-normal distribution. We find
\begin{equation}
r_{\rm{c}} = (0.51 \pm 0.24)\, \rm{mm}.
\end{equation} 
Three typical spirals constructed by considering a constant value of $r_{\rm{c}}$ equal to its expectation value and $\theta_0$ varying by one standard deviation around its expectation value are shown in Fig.\ref{fig05}B. Those three spirals are generated from the same expanding bubble of reagent having a radius equals to $40\, r_{\rm{c}}$. Consequently, they all have the same couple of radii of curvature at their end points. An animation showing the growth of these spirals can be found as Supporting Information (see Movie S2). Movie S3 shows a qualititative comparison between the growth a spiral observed experimentally and a spiral obtained from the geometrical model.

Finally, Fig.\ref{fig06} shows the distribution of the maximal value, $\theta_{\rm{max}}$, of the polar angle characterizing each analyzed spiral. As explained in the main text, we choose $\theta_{\rm{max}} > 2.44$ $(140^{\circ})$ such that the maximal value of $\theta /\theta_0$ for each spiral is large enough to obtain a relevant comparison with the model.

\begin{figure*}[h]
\begin{center}
\includegraphics[width=\textwidth]{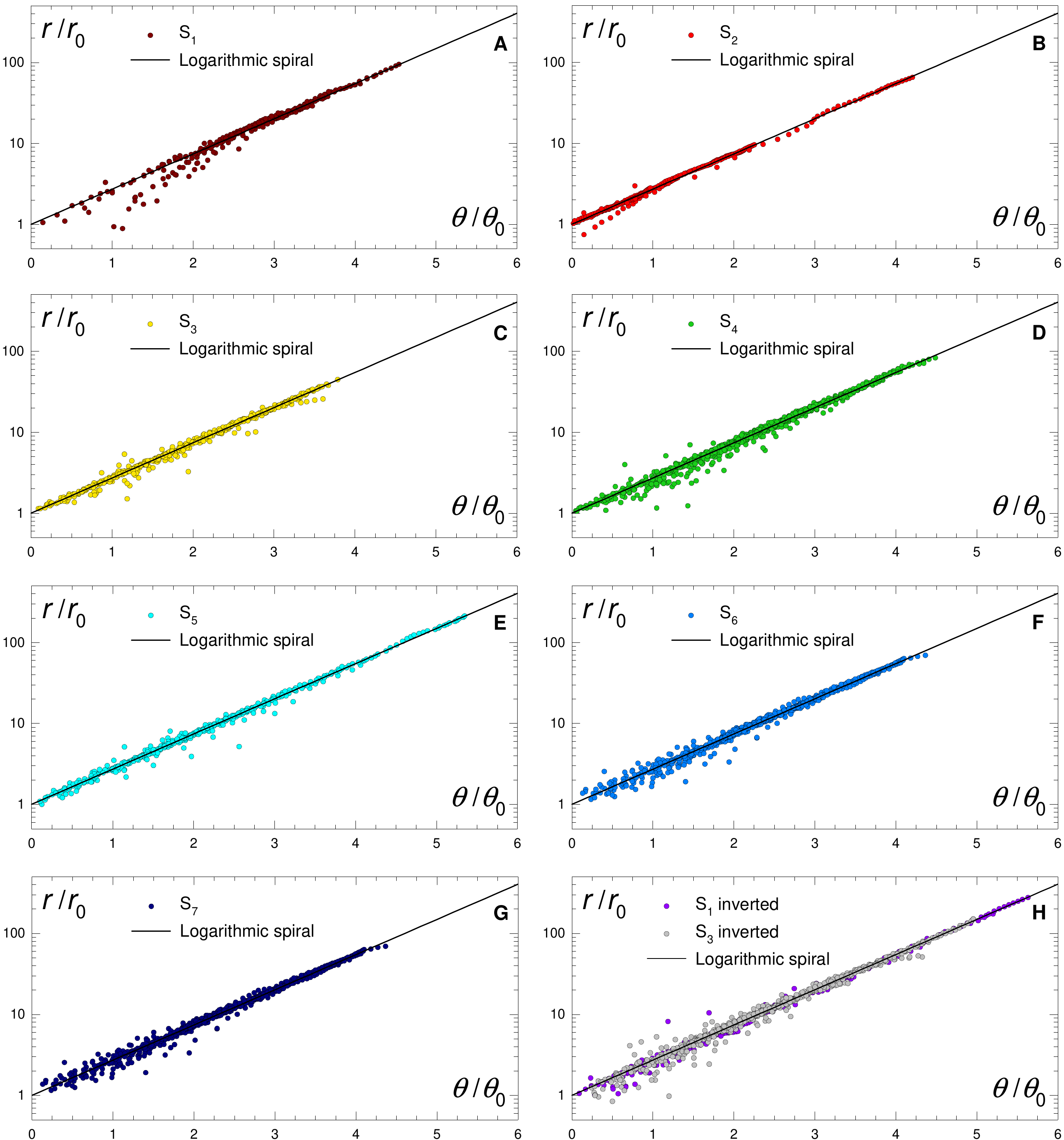} 
\end{center}
\caption{{\bf A-G}, Evolution of the rescaled spiral radii, $r/r_0$ as a function of the rescaled polar angle, $\theta/\theta_0$, for each category $S_i$ identified in Fig.1 of the main text. {\bf H}, Evolution of the rescaled spiral radii, $r/r_0$ as a function of the rescaled polar angle, $\theta/\theta_0$, for the inverted case, where sodium silicate is injected into cobalt chloride, corresponding to the sectors $S_1$ and $S_3$.}
\label{fig03}
\end{figure*}

\begin{figure*}
\begin{center}
\includegraphics[width=\textwidth]{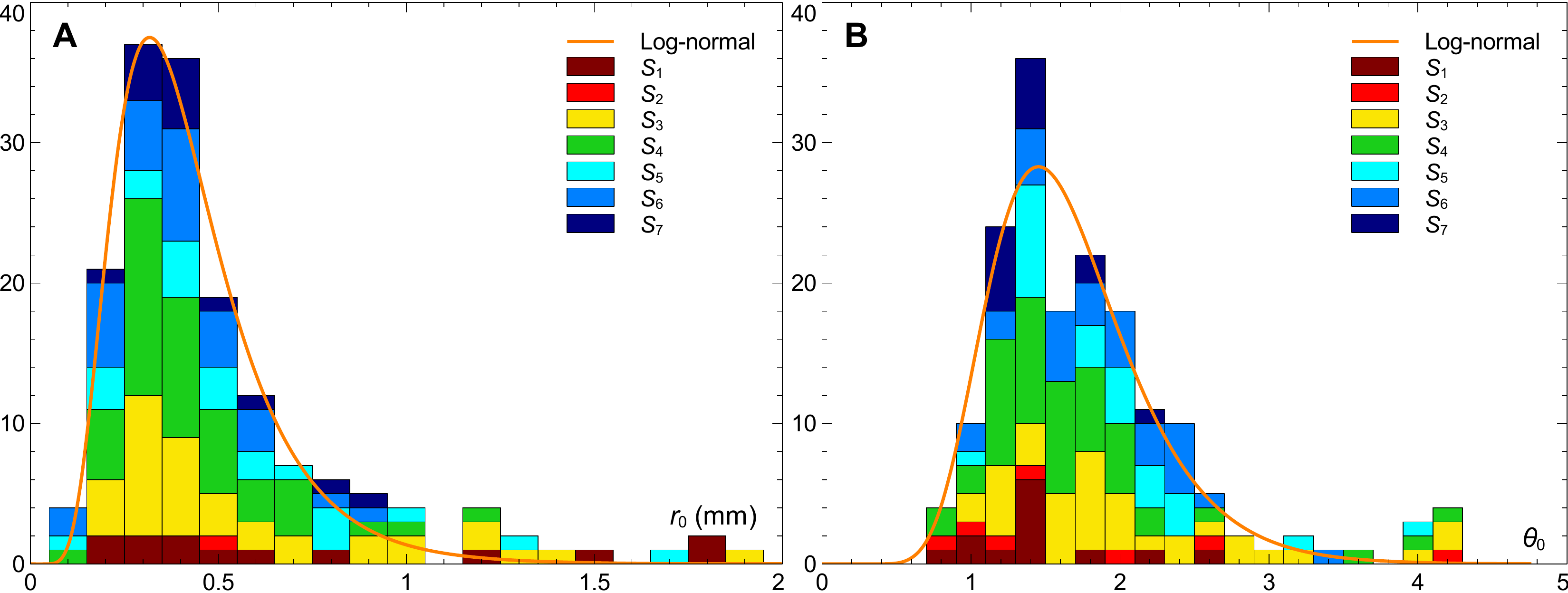} 
\end{center}
\caption{{\bf A,} Distribution of $r_0$ and {\bf B,} distribution of  $\theta_0$ for all analyzed spirals. The contributions of each sector $S_i$ is indicated. The log-normal distributions (\ref{log-normal}) are characterized by $\mu=-0.97$ and $s=0.45$ for $r_0$ and $\mu=0.47$ and $s=0.31$ for $\theta_0$.} 
\label{fig04}
\end{figure*}

\begin{figure*}
\begin{center}
\includegraphics[width=\textwidth]{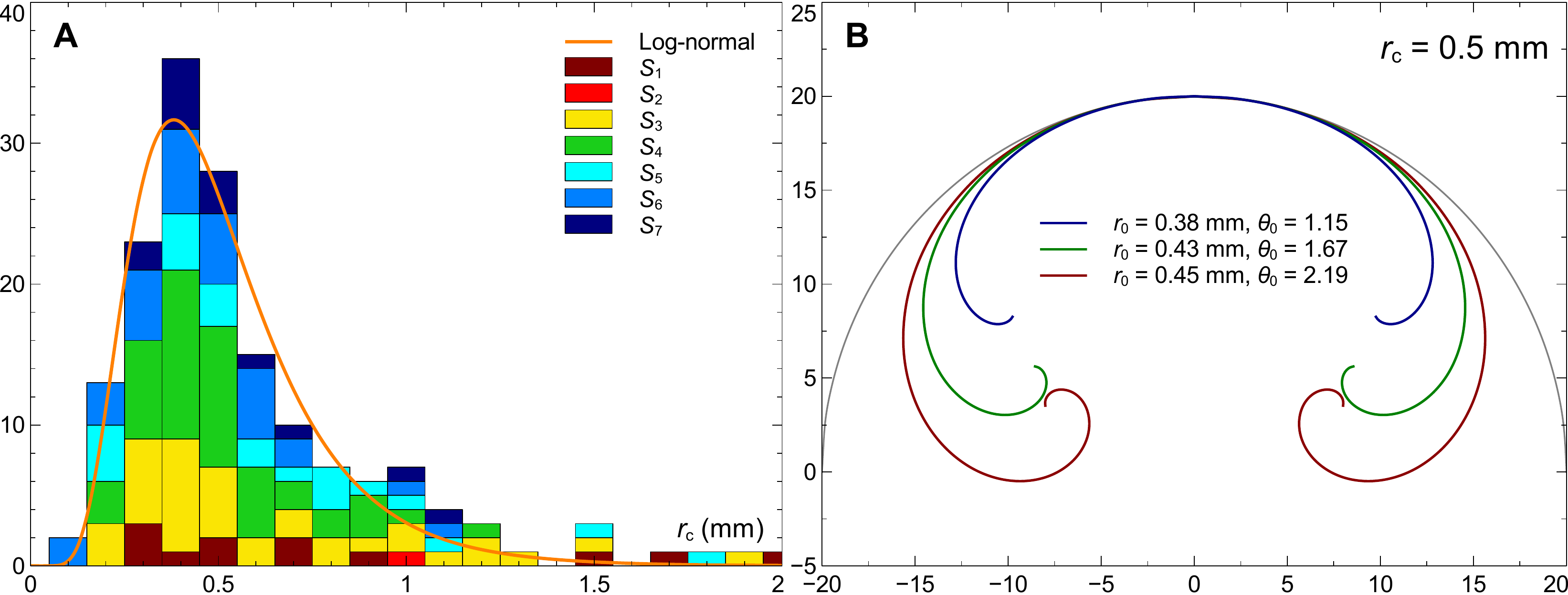} 
\end{center}
\caption{{\bf A,} Distributions of $r_{\rm{c}}$ and for all analyzed spirals. The contributions of each sector $S_i$ is indicated. The log-normal distribution (\ref{log-normal}) is characterized by $\mu=-0.77$ and $s=0.47$. {\bf B,} Three typical spirals constructed by considering a constant value of $r_{\rm{c}}$ equal to its expectation value and $\theta_0$ varying by one standard deviation around its expectation value. The spirals are generated from the same expanding bubble (in gray) of reagent having a radius equals to $40\, r_{\rm{c}}$. Consequently, they all have the same couple of radius of curvature at their end points. The graph graduations are in millimeters.} 
\label{fig05}
\end{figure*}

\begin{figure*}
\begin{center}
\includegraphics[width=0.5\textwidth]{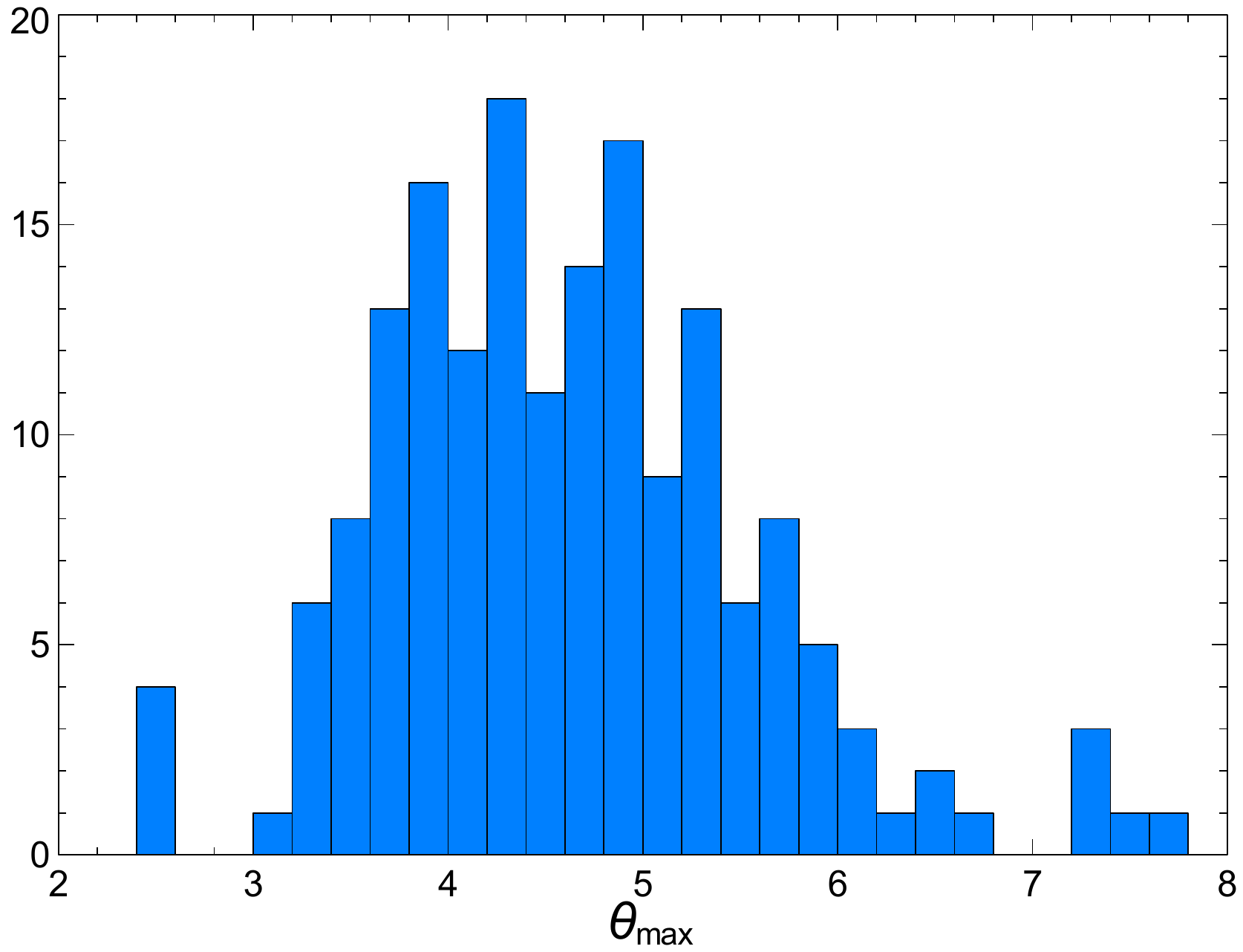} 
\end{center}
\caption{Distribution of $\theta_{\rm{max}}$ defined as the maximal value of the polar angle describing each analyzed spiral.} 
\label{fig06}
\end{figure*}

\end{article}

\end{document}